\documentclass[oneside,a4paper]{article}

\usepackage{hyperref}
\usepackage{amssymb}
\usepackage{csquotes}
\usepackage{booktabs}
\usepackage{graphicx}
\usepackage{amsmath}
\graphicspath{ {./ }}
\usepackage{authblk}
\providecommand{\keywords}[1]{\textbf{\textit{Keywords:}} #1}

\title{A survey on making skylines more flexible}
\author{Cem Cebeci}
\affil{Politecnico di Milano\\
Milan, Italy\\
\href{mailto:cem.cebeci@mail.polimi.it}{cem.cebeci@mail.polimi.it} }
\date{\today}

\begin{document}
\maketitle
\begin{abstract}
    Top-$k$ queries and skylines are the two most common approaches to finding the most interesting entries in a homogeneous multi-dimensional dataset.
    However, both of these strategies have some shortcomings.
    Top-$k$ queries are very challenging to specify precisely and skylines are not customizable to specific scenarios, on top of having unpredictable output cardinalities.
    We describe some alternative methods aimed at adressing the shortcomings of top-$k$ queries and sklyines and compare all approaches to illustrate which of the desired properties each of them possesses.
\end{abstract}


\section{Introduction}
Selecting the most interesting entries from a multi-dimensional dataset is an important operation in today's recommender systems \cite{skylineRecommendation}\cite{topkRecommendation}. 
In order to perform this task, the two most commonly used approaches are top-$k$ queries \cite{topksurvey} and skylines \cite{SkylineOperator}.

Given a multi-dimensional dataset $R$ and a scoring function $f: R \rightarrow \mathbb{R}$, the top-$k$ query on $R$ with $f$ returns the $k$ tuples $r_1,...,r_k \in R$
for which $f(r_i)$ is smaller than any tuple not included in the query result. Top-$k$ queries are expressive and precise and we have fast algorithms \cite{fagin}\cite{threshold} to compute them.
However, defining "most interesting" as a function of dataset attributes proves to be a significant challenge, even though techniques such as crowdsourcing \cite{crowdsourcing} 
provide some tangible results. Additionally, query results may differ significantly with slight alterations on $f$ depending on the dataset, which poses a considerable problem given
the natural inaccuracy we have in determining $f$.

On the other hand, a skyline query returns the tuples in a dataset that are not Pareto dominated. In other words, it picks the tuples $r$ for which there are no other tuples $t$
such that $t$ is no worse than $r$ for all attributes and better than $r$ for at least one attribute. It's worth noting that tuples in the skyline are the top-1 result for some
scoring function. The notion of a skyline can be extended to a $k$-skyband, the tuples that have less than $k$ tuples Pareto dominating them, or equivalently, top-$k$ results
for some scoring function.

Skylines do not suffer from the same challenge as top-$k$ queries do since they require no scoring function to be specified. However, due to the same reason, they are not in any way customizable.
There is no way to alter a skyline query to fit a user's preferences, which is crucial in recommender systems. Another shortcoming of skyline queries is that their output cardinality is unbounded.
In fact, for datasets with high dimensionality, the skyline may comprise the whole dataset \cite{ORD/ORU}, which renders the skyline operator useless. In general, a high number of results is undesirable
in skyline queries.

To address the shortcomings of skylines and top-$k$ queries, multiple alternative operators have been introduced.
In this paper, we describe some of those proposed solutions and investigate how they relate to each other in Section \ref{proposedSolutions}.
We compare the different approaches in Section \ref{comparison} and conclude in Section \ref{conclusion}.

\section{Proposed Solutions}\label{proposedSolutions}
\subsection{Flexible Skylines}\label{flexibleSkylines}
As mentioned before, one of the major shortcomings of skylines is customization.
The skyline of a dataset does not depend on any kind of parameter, only on the dataset itself.
As a result, the skyline query can not be customized in any way.
$F$-skylines are a generalization of the skyline that introduces a parameter to the query while compromising very little from the simplicity of skyline queries \cite{fskylines}{}.

Before we define what an $F$-skyline is, let us define $F$-domination. A tuple $r_1$ $F$-dominates another tuple $r_2$ if and only if $r_1$ is at least as good as $r_2$ for all scoring functions in $F$ and better than $r_2$ for at least one scoring function. In formal terms:

\[r_1 \prec r_2 \iff \forall f \in F . \  f(r_1) \le f(r_2) \land \exists f \in F . f(r_1) < f(r_2)\]

If $F$ is the set of all monotonic functions, $F$-domination reduces to Pareto domination.

Given a dataset $R$ and a set of monotone scoring functions $F$, the two flexible skyline operators are
defined as follows:
\[ND(R, F) := \{r \in R \ \vert\ \nexists t \in R.\ t \prec_F r \}\]
\[PO(R, F) := \{r \in R \ \vert\ \exists f \in F,\forall r' \in r.\ f(r) \le f(r')\}\]
The first operator, $ND$, denotes all non $F$-dominated tuples of $R$ and the second operator, $PO$,
denotes the tuples in $R$ that are optimal for at least one of the functions in $F$. When $F$ is the set of all monotonic
functions, both $ND(R,F)$ and $PO(R,F)$ converge to the skyline of $R$. However, in the general case, they are not
necessarily equal \cite{fskylines}.

$ND$ and $PO$ are both customizable since they have an additional parameter, $F$, that enables customized queries.
In addition, they partially address the other shortcoming of skylines, they lower the output cardinality for non-universal
choices of $F$. One can show $PO(R,F) \subseteq ND(R,F) \subseteq SKY(R)$ for all choices of $R$ and $F$. Thus guaranteeing
neither $ND$, nor $PO$ will have a greater cardinality than the skyline. Experimental results \cite{fskylines}
depict that the cardinality of $ND$ is dramatically lower than the skyline even with only a few constraints on $F$ and the 
cardinality of $PO$ is even lower.

Customizability and control over result cardinality, the two properties flexible skylines provide, are already available
in top-$k$ queries, although flexible skylines do not require precisely defined scoring functions as top-$k$ queries do.
As an example, consider a simplified universe of functions comprising only weighted sums over the attributes with normalized
weights, functions of the form $f(r) = \sum{w_ir_i}$ where $r_i$ denotes an attribute in the schema of $R$.
A top-$k$ query would need to specify a precise scoring function (or equivalently, a precise vector $w$) but a flexible skyline
allows less precise queries such as $F=\{f(r) =w_1r_1+w_2r_2\ \vert\  w_1 > 2w_2\}$. Such constraints are significantly easier
to obtain \cite{pairwise}. For instance, if a user prefers the tuple $(2,3)$ to $(1,6)$, we can argue:
\[2w_1 + 3w_2 < w_1 + 6w_2,\  i.e.,\]
\[w_1 < 3w_2\]

Flexible skylines merge the idea of scoring functions of top-$k$ queries with traditional skylines.
This is further illustrated by the fact that choosing $F$ to be the set of all monotonic
functions reduces both operators to the traditional skyline but choosing it to contain a single monotonic function reduces
them to the top-1 query result. By combining the two approaches, flexible skylines are both easy to specify and customizable
at the same time.

The two flexible operators are discussed only for skylines here for brevity, but \cite{UTK} and \cite{FSA} extend this
notion to $k$-skybands to provide a more generalized framework.

\subsection{Output Size Specified (OSS) operators}
Another solution approach \cite{ORD/ORU} to the same problem introduces the notion of $\rho$-dominance and two operators that utilize 
this relation to provide types of queries that are not as precise as top-$k$ queries but still have query customization and smaller output cardinality
than the skyline. In fact, these operators allow specifying the cardinality of their output as a parameter.

These operators only use weighted sums over attributes with normalized preferences as scoring functions. They define a new kind of dominance, much like 
$F$-dominance. Given an estimate preference vector $w$ and tuples $r_1, r_2$, $r_1$ $\rho$-dominates $r_2$ if the following conditions are satisfied:

\begin{center}
    \begin{enumerate}
        \item $\forall v \in \Delta^{d-1}.\  |w-v| \le \rho \implies \sum_{i}{v_i {r_1}_i \le \sum_{i}{v_i {r_2}_i}}$
        \item $\exists v \in \Delta^{d-1}.\  |w-v| \le \rho \land \sum_{i}{v_i {r_1}_i < \sum_{i}{v_i {r_2}_i}}$
    \end{enumerate} 
\end{center}


In other words, they relax the normalized preference vector $w$ up to a distance of $\rho$ and check dominance for every preference vector inside
the resulting hypersphere. If a $r_1$ performs at least as good as $r_2$ for all vectors and better than $r_2$ for at least one, $r_1$ $\rho$-dominates
$r_2$, much like the concept of domination in flexible skylines.

The first output size specified operator is called $ORD$, it is \textbf{O}utput size specified, has \textbf{r}elaxed input and is \textbf{d}ominance
oriented. Given an estimate preference vector $w$, an output size $m$ and a dataset $R$, $ORD(R,w,m)$ denotes $m$ tuples in $R$ that are not $\rho$-dominated
for the minimum value of $\rho$ that allows $m$ tuples.

The second output size specified operator is $ORU$, again because it is \textbf{O}utput size specified, has \textbf{r}elaxed input and it is
\textbf{u}tility-oriented. $ORU(R,w,m)$ denotes $m$ tuples in $R$ that are optimal for at least one preference vector with a maximum distance of $\rho$ from
$w$.

Both of these operators can be further generalized by relaxing $ORD$ to contain tuples that are dominated by fewer than
$k$ others and relaxing $ORU$ to contain tuples that are in the top-$k$ result for at least one vector.

This approach is very similar to flexible skylines. In fact, for any preference vector $v$ and distance $\rho$, weighted sums with
weight vectors inside the hypersphere centered around $v$ with radius $\rho$ specifies a set of functions $F$. In fact, for any dataset $R$, 
non-$\rho$-dominated tuples in $R$ are precisely $ND(R,F)$ and tuples that are optimal for at least one vector are precisely $PO(R,F)$.

\subsection{UTK queries}
Another parallel method that addresses the problem of relaxing the input preference vector is \textbf{u}ncertain \textbf{t}op-\textbf{$K$} (UTK) queries \cite{UTK}.
Similar to OSS operators, these queries only consider weighted sums over the attributes as scoring functions. Likewise, each UTK query operates on a set of possible
preference vectors rather than a single vector and these possible preference vectors form a convex polytope in the $d-1$ dimensional simplex.
Instead of specifying the region of possible vectors via a central vector and a radius, UTK queries allow the region to be specified as a parameter on its own.

\cite{UTK} defines two different UTK operators, $UTK_1$ and $UTK_2$. Given a $d$ dimensional dataset $R$, a desired output cardinality $k$ and a
preference region $P \subseteq \Delta ^{d-1}$, $UTK_1(k,R,P)$ contains the tuples in $R$ that are in the top-$k$ result for a preference vector in $P$ and
$UTK_2(k,R,P)$ partitions $P$ so that vectors in each partition produce the same top-$k$ result and labels the partitions with the top-$k$ results they correspond to.

UTK queries can express a larger family of scoring functions than OSS operators since every $d-1$ dimensional hypersphere is a $d-1$ dimensional polytope but the converse
is not true. However, flexible skylines can express an even larger family of scoring functions since UTK is still limited to functions that are linear in the tuple attributes.

The sets of tuples computed by $UTK_1$ queries correspond to the tuples computed by $PO$ and $ORU$ operators for matching inputs. Since $UTK$ queries only check optimality and
not domination, $ND$ and $ORD$ can not be computed by $UTK$ queries.

\subsection{$\epsilon$-skylines}
$\epsilon$-skylines \cite{epsilonSkylinkes} is one of the earlier approaches to addressing the shortcomings of top-$k$ and skyline queries.
Similar to OSS operators, $\epsilon$-skylines only consider weighted sums as scoring functions. They use the notion of $\epsilon$-dominance,
which, given a dataset $R$, a normalized preference vector $w$ and a constant $\epsilon \in [-1,1]$ is defined as:
\[
    r_1 \prec_\epsilon r_2 \iff (\forall i. w_i{r_1}_i \le w_i{r_2}_i + \epsilon)\  \land\ (\exists i .{r_1}_i < {r_2}_i)
\]
it is a relaxed version of Pareto dominance where the dominant tuple is allowed to be worse in some attributes up to an additive constant $\epsilon$.
In fact, when we pick $\epsilon=0$, $\epsilon$-dominance reduces to Pareto dominance.

The $\epsilon$ skyline of a dataset for a preference vector $w$ and constant $\epsilon$ is simply the set of non-$\epsilon$-dominated
tuples. Unlike traditional skylines, $\epsilon$-skylines are customizable in both preferences and output cardinality. Since 
$\epsilon$ is scaled with the weights $w_i$ while checking dominance, changing $w$ enables the operator to respond to the preferences
of a user. The impact of $\epsilon$ on dominance decreases as the weight $w_i$ of an attribute increases.

Regarding the output cardinality, $\epsilon$-skylines do not provide a precise number but allow controlling the cardinality via changing
$\epsilon$. For positive $\epsilon$, every tuple's $\epsilon$-dominance region is larger than their Pareto dominance region, resulting in a smaller number of
non-dominated tuples. Conversely, negative $\epsilon$ shrink the $\epsilon$-dominance regions and result in a larger number of non-dominated tuples.
In fact, for datasets with traditional skylines that contain at least two tuples, $\epsilon=-1$ produces the whole dataset and $\epsilon=1$ produces the
empty set\cite{epsilonSkylinkes}.

\subsection{Representative Skylines}
The approaches described so far have focused on customizing the query somehow to include user preferences. Representative skylines
\cite{selectingStars}\cite{distanceRepresentative} take an orthogonal approach, they reduce the cardinality of the skyline directly, by
picking $k$ representative tuples from the skyline. The original implementation \cite{selectingStars} picks the tuples that maximize
the number of non-skyline tuples dominated by the picked tuples. A variant \cite{distanceRepresentative} of this implementation picks the tuples that minimize the
maximum distance between a non-picked skyline tuple and any picked tuple.

\begin{figure}[h]
    \centering
    \includegraphics[width=6cm]{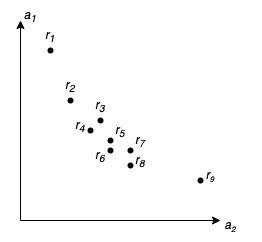}
    \caption{An example skyline illustrating drawbacks}
    \label{fig:rep}
\end{figure}

Both versions have some drawbacks \cite{distanceRepresentative}. Consider the example in Figure \ref{fig:rep}, the skyline of the given dataset is $\{r_1, r_2, r_4, r_6, r_8, r_9\}$. If we 
optimize the number of dominated non-skyline tuples and pick $k=3$, we get $\{r_4, r_6, r_8\}$. These two tuples are very similar in the trade-off they offer between attributes
$a_1$ and $a_2$. This subset fails to represent the overall skyline since it does not include any other types of trade-off in the skyline and focuses on a small cluster of tuples. On the other hand,
minimizing the maximum distance gives $\{r_1, r_4, r_9\}$. This selection presents other trade-offs much better, but it fails to capture that most of the tuples in the skyline have a trade-off similar To
that of $r_4$.

Representative skylines have the benefit of requiring no user input. They provide a way to limit output cardinality, with different
diversification techniques, without compromising the simplicity of skyline queries. On the other hand, since they admit no input other than
the dataset itself, they are not customizable save for the output cardinality.

\section{Comparison}\label{comparison}
In this section, we discuss how the various proposed solutions compare to each other, as well as traditional top-$k$ queries and skylines.
We analyze which problems each solution can address by inspecting four properties: input flexibility, customizability, control over cardinality and
ranked output. These properties will be discussed one by one in the following subsections. This discussion is summarized in Table \ref{table:comparison}.

\begin{table}[h]
    \begin{tabular}{|l|l|l|l|l|}
    \hline
                            & Input Flexibility & Customizability & Cardinality Control & Ranked Output \\ \hline
    Top-$k$ queries         & No                & Yes             & Yes                 & Yes           \\ \hline
    Traditional Skylines    & Yes               & No              & No                  & No            \\ \hline
    Flexible Skylines       & Yes               & Yes             & Partial             & No            \\ \hline
    OSS Operators           & Yes               & Partial         & Yes                 & No            \\ \hline
    UTK Queries             & Yes               & Partial         & Yes                 & No            \\ \hline
    $\epsilon$-skylines     & Yes               & Partial         & Partial             & No            \\ \hline
    Representative Skylines & Yes               & No              & Yes                 & Yes           \\ \hline
    \end{tabular}
    \caption{Properties of different approaches}
    \label{table:comparison}
\end{table}

\subsection{Input flexibility}
A query's input is said to be flexible if it does not need to be precise and if small changes in the input do not cause significant changes in the output.
Traditional skylines and representative skylines are the obvious superior choices for input flexibility since they have no input but the dataset itself. Thus,
no additional parameter needs to be defined to make these queries. $\epsilon$-skylines are a bit less flexible, they require a value for $\epsilon$ and a weight vector but
the weight vector does not affect the output as directly as it does in some other approaches. OSS operators, flexible skylines and UTK queries are somewhat less flexible because they require some specifics about
either the estimate preference weights or the family of scoring functions. Finally, top-$k$ queries are easily the least flexible of the queries we consider, they require a
precisely defined scoring function which affects the output dramatically.

\subsection{Customizability}
A query is said to be customizable if it can adapt to the specific requirements of a user or a scenario.
Traditional skylines are not customizable in any way, they simply return the same set of tuples for every scenario. Representative skylines, however, are mildly customizable
since the output cardinality can be controlled. $\epsilon$-skylines, OSS skylines and UTK queries have noticeably higher customizability since they allow weight vectors to represent preferences
over the attributes, though they only allow preferences to be expressed in the form of weighted sums. Top-$k$ queries and flexible skylines are fully customizable, they can be
configured to return any tuple in the skyline.

\subsection{Control over cardinality}
The approaches are clearly divided into three groups on this property: Traditional skyline queries offer no control over the output cardinality.
$\epsilon$ skylines and flexible skylines provide methods to increase or decrease the cardinality but offer no precise control. Lastly, top-$k$
queries, representative skylines, UTK queries and OSS operators have a parameter to specify the output cardinality.

\subsection{Ranked output}
Top-$k$ queries and representative skylines provide a ranking between the tuples they return. None of the other approaches have this property.

\section{Conclusion}\label{conclusion}
In this paper, we mentioned the commonly used strategies to select the most interesting tuples in homogeneous multi-dimensional datasets together with their well-known shortcomings.
Then, we described some strategies that aim to address the shortcomings of the conventional methods:
Flexible skylines, output size specified operators, UTK queries, $\epsilon$-skylines and representative skylines.
Finally, we compared these proposed solutions with each other and the traditional strategies, pointing out which issues each one can address.

Future work on flexible skylines could include estimating the output cardinality or limiting it by construction similar to what output size specified operators do. 
Discovering other families of scoring functions that are representative of users' preferences and can be efficiently computed by flexible skylines could also be worthwhile research.
Ranking the items contained in flexible skylines and computing the skyline tuple-by-tuple could be other directions to improve on this notion.

\bibliographystyle{plain}
\bibliography{refs.bib}

\begin{thebibliography}{10}

\bibitem{SkylineOperator}
Stephan B{\"o}rzs{\"o}nyi, Donald Kossmann, and Konrad Stocker.
\newblock The skyline operator.
\newblock {\em Proceedings 17th International Conference on Data Engineering},
  pages 421--430, 2001.

\bibitem{FSA}
Paolo Ciaccia and Davide Martinenghi.
\newblock Fa + ta {<} fsa: Flexible score aggregation.
\newblock CIKM '18, page 57–66, New York, NY, USA, 2018. Association For
  Computing Machinery.

\bibitem{fskylines}
Paolo Ciaccia and Davide Martinenghi.
\newblock Flexible skylines: Dominance for arbitrary sets of monotone
  functions.
\newblock {\em ACM Trans. Database Syst.}, 45(4), dec 2020.

\bibitem{crowdsourcing}
Eleonora Ciceri, Piero Fraternali, Davide Martinenghi, and Marco Tagliasacchi.
\newblock Crowdsourcing for top-k query processing over uncertain data.
\newblock {\em IEEE Transactions on Knowledge and Data Engineering},
  28(1):41--53, 2016.

\bibitem{fagin}
Ronald Fagin.
\newblock Combining fuzzy information from multiple systems (extended
  abstract).
\newblock In {\em Proceedings of the Fifteenth ACM SIGACT-SIGMOD-SIGART
  Symposium on Principles of Database Systems}, PODS '96, page 216–226, New
  York, NY, USA, 1996. Association for Computing Machinery.

\bibitem{threshold}
Ronald Fagin, Amnon Lotem, and Moni Naor.
\newblock Optimal aggregation algorithms for middleware.
\newblock In {\em Proceedings of the Twentieth ACM SIGMOD-SIGACT-SIGART
  Symposium on Principles of Database Systems}, PODS '01, page 102–113, New
  York, NY, USA, 2001. Association for Computing Machinery.

\bibitem{topksurvey}
Ihab~F. Ilyas, George Beskales, and Mohamed~A. Soliman.
\newblock A survey of top-<i>k</i> query processing techniques in relational
  database systems.
\newblock {\em ACM Comput. Surv.}, 40(4), oct 2008.

\bibitem{skylineRecommendation}
Shuhei Kishida, Seiji Ueda, Atsushi Keyaki, and Jun Miyazaki.
\newblock Skyline-based recommendation considering user preferences.
\newblock In Lei Chen, Christian~S. Jensen, Cyrus Shahabi, Xiaochun Yang, and
  Xiang Lian, editors, {\em Web and Big Data}, pages 133--141, Cham, 2017.
  Springer International Publishing.

\bibitem{selectingStars}
Xuemin Lin, Yidong Yuan, Qing Zhang, and Ying Zhang.
\newblock Selecting stars: The k most representative skyline operator.
\newblock In {\em 2007 IEEE 23rd International Conference on Data Engineering},
  pages 86--95, 2007.

\bibitem{ORD/ORU}
Kyriakos Mouratidis, Keming Li, and Bo~Tang.
\newblock {\em Marrying Top-k with Skyline Queries: Relaxing the Preference
  Input While Producing Output of Controllable Size}, page 1317–1330.
\newblock Association for Computing Machinery, New York, NY, USA, 2021.

\bibitem{UTK}
Kyriakos Mouratidis and Bo~Tang.
\newblock Exact processing of uncertain top-k queries in multi-criteria
  settings.
\newblock {\em Proc. VLDB Endow.}, 11(8):866–879, apr 2018.

\bibitem{pairwise}
Li~Qian, Jinyang Gao, and H.~V. Jagadish.
\newblock Learning user preferences by adaptive pairwise comparison.
\newblock {\em Proc. VLDB Endow.}, 8(11):1322–1333, jul 2015.

\bibitem{distanceRepresentative}
Yufei Tao, Ling Ding, Xuemin Lin, and Jian Pei.
\newblock Distance-based representative skyline.
\newblock In {\em 2009 IEEE 25th International Conference on Data Engineering},
  pages 892--903, 2009.

\bibitem{epsilonSkylinkes}
Tian Xia, Donghui Zhang, and Yufei Tao.
\newblock On skylining with flexible dominance relation.
\newblock In {\em 2008 IEEE 24th International Conference on Data Engineering},
  pages 1397--1399, 2008.

\bibitem{topkRecommendation}
Xiwang Yang, Harald Steck, Yang Guo, and Yong Liu.
\newblock On top-k recommendation using social networks.
\newblock In {\em Proceedings of the Sixth ACM Conference on Recommender
  Systems}, RecSys '12, page 67–74, New York, NY, USA, 2012. Association for
  Computing Machinery.

\end{thebibliography}
\end{document}